\documentclass[a4paper]{ar-1col}

\usepackage{amsmath,amssymb}

\newcommand{\be}{\begin{equation}}
\newcommand{\ee}{\end{equation}}

\jname{Ann. Rev. Nuc. Part. Sci.}
\jvol{66}
\jyear{2016}
\doi{10.1146/(DOI)}

\begin{document}

\markboth{Fern\'andez \& Metzger}{EM Signatures of NS Mergers in the aLIGO Era}

\title{Electromagnetic Signatures of Neutron Star Mergers in the Advanced LIGO Era}

\author{Rodrigo Fern\'andez$^{1,2}$ and Brian D. Metzger$^3$
\affil{$^1$Department of Physics, University of California, Berkeley, CA, 94720, USA}
\affil{$^2$Department of Astronomy \& Theoretical Astrophysics Center, University of California, Berkeley, CA, 94720, USA}
\affil{$^3$Columbia Astrophysics Laboratory, Columbia University, New York, NY, 10027, USA}}

\firstpagenote{First page note to print below DOI/copyright line.}

\begin{abstract}
The mergers of binaries containing neutron stars and stellar-mass black holes
are the most promising sources for direct detection in gravitational waves by
the interferometers Advanced LIGO and Virgo over the next few years.  The
concurrent detection of electromagnetic emission from these events would
greatly enhance the scientific return of these discoveries.  Here we review
the state of the art in modeling the electromagnetic signal of neutron star
binary mergers across different phases of the merger and multiple wavelengths.
We focus on those observables which provide the most sensitive diagnostics of
the merger physics and the contribution to the synthesis of rapid neutron
capture ($r$-process) elements in the Galaxy.  We also outline expected future
developments on the observational and theoretical sides of this rapidly
evolving field.
\end{abstract}

\maketitle

\tableofcontents

\section{INTRODUCTION}

Observations of the Hulse-Taylor binary pulsar \cite{Taylor&Weisberg89} and of
the double pulsar J0737-3039 \cite{Lyne+04} have proven the existence of gravitational 
waves (GWs).  However, their direct detection
remains elusive.  This situation is anticipated to change soon, as a network of
new GW observatories, including Advanced LIGO \cite{Harry+10} (herafter aLIGO),
Advanced Virgo \cite{Degallaix+13}, and the Japanese
kryogenic detector KAGRA \cite{Somiya12}, begin taking data over the next few
years.  The first detection of GWs will test a key prediction of General
Relativity (GR) and mark the beginning of a new field: GW Astronomy.

Among the most likely sources to be detected in this initial phase are coalescing
binaries containing two neutron stars (NS), hereafter NSNS mergers, or a NS
and a black hole (BH), hereafter NSBH mergers \cite{Abadie+10}\footnote{We do not
consider double BH binaries here, see e.g. \cite{lehner2014} for a review.}.
These detectors are most sensitive to 
GWs from the late stages of the binary inspiral. 
At design sensitivity, aLIGO will be capable of detecting NSNS mergers out to 
an orientation-averaged distance $\sim 200$~Mpc, and NSBH mergers to a distance $\approx 2-3$
times larger \cite{Abadie+10}.
The planned network of GW interferometers can narrow
down the location of a source using triangulation, but depending on the signal
to noise ratio the final uncertainty can still be tens or hundreds of square degrees, 
translating into large uncertainties in the parameters inferred
from the GW waveform [e.g., \cite{singer2014}].

Astronomers have refined the measurement of electromagnetic (EM) waves for
centuries, and optical, near-infrared (IR), and many radio observatories have
exquisite localization precision ($\sim$ arcseconds) compared with GW
observatories.  For this reason alone, the detection of an EM counterpart to a
GW source would greatly improve the quality of the information available from
the GW data.  It could also provide complementary information, such as the
energetics and host galaxy of the event, and when (or whether) an event horizon
forms.  Obtaining an accurate source position is multiplicative, as it enables
a much larger range of electromagnetic facilities (often more sensitive, but
with narrower fields of view,
e.g. the Hubble Space Telescope)
to obtain complementary observations.  
 
Due to their transient nature, discovering the EM counterparts of
NSNS/NSBH mergers requires follow-up observations with time-sensitive
facilities.  NASA's  {\it Swift}  and {\it Fermi} satellites provide nearly
continuous coverage of the sky at hard X-ray and gamma-ray wavelengths.
Several optical/IR transient surveys have been in operation over the last
several years, with more coming online in the next several years, culminating
in the Large Synoptic Survey Telescope \cite{ivezic2008lsst}.  Wide-field
radio arrays, such as LOFAR \cite{VanHaarlem+13}, provide nearly continuous
coverage of the northern hemisphere sky in the hundreds of MHz radio band.

NSNS/NSBH mergers also represent an important topic in Nuclear Astrophysics.
The neutron star equation of state (EOS) plays an
important role in the GW signal, both during the late inspiral phase and in the
fate of the post-merger remnant.  The ejecta from NS mergers are an 
astrophysical source of rapid neutron-capture ($r$-process) nuclei, the origin
of which has remained a mystery for almost 70 years \cite{BBFH57,cameron1957}.  The short-lived,
neutron-rich nuclei produced during the $r$-process serve as probes of the
nuclear force in asymmetric conditions and of the limits of nuclear stability
\cite{nuc_lrp_2015}.  Significant efforts are underway to improve experimental
capabilities to measure the masses and lifetimes of these nuclei, including the
Facility for Rare Isotope Beams \cite{frib2010}. 

\begin{figure}
\includegraphics*[width=5.5in]{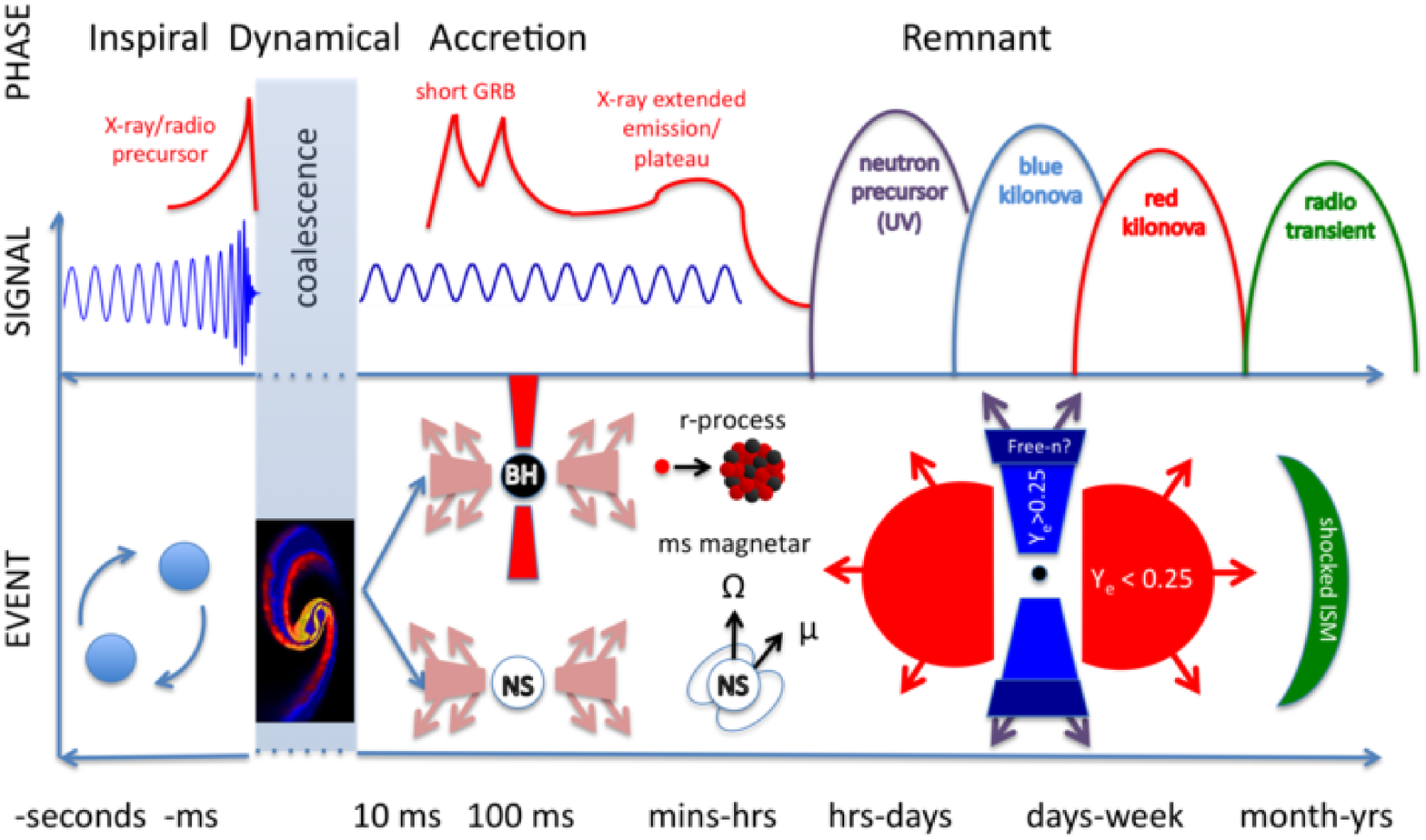}
\caption{Phases of a neutron star merger as a function of time, showing
the associated observational signatures and underlying physical phenomena.
Coalescence inset courtesy of D. Price and S. Rosswog [see also \cite{Price&Rosswog06}].}
\label{f:merger_phases}
\end{figure}

This review summarizes the state-of-the art in the predicted EM emission from
NSNS and NSBH mergers. The structure of the paper follows the time evolution of the
merger event, as shown in {\bf Figure~\ref{f:merger_phases}}. Future issues
are summarized at the end of the article.
Due to space constraints, we
are not able to cover and/or cite all the work in the field;
we refer the reader to 
other excellent reviews on this topic \cite{Lee&RamirezRuiz07,Berger14,Rosswog15}.

\section{INSPIRAL PHASE}
\label{s:inspiral}

\subsection{Formation and Rates}

The GW detection rates of NSNS/NSBH mergers are uncertain due to their poorly constrained volumetric rate in the local universe.  Ten NSNS binary systems are currently known in our Galaxy, of which 6 will coalesce within a Hubble time [Table 2 of \cite{swiggum+15}].  This population provides an empirically-calibrated estimate of the NSNS rate per stellar mass \cite{Kalogera+04}, which when translated to the Galaxy population as a whole results in an aLIGO/Virgo detection rate of 3-18 yr$^{-1}$ \cite{Kim+15}.  This estimate is, however, subject to several uncertainties, such as the faint end of the pulsar luminosity function and small number statistics (the rates are dominated by J0737-3039, which has a particularly short inspiral time).  

No NSBH binaries are currently known, so the empirical method cannot be applied to estimate their rates.   An upper limit is set by the galactic formation rate of High Mass X-ray Binaries (HMXBs) of $\sim 10^{-3}$ yr$^{-1}$, which when combined with the fraction of those which pass through a bright X-ray phase of a few percent, results in a rate of less than a few dozen NSBH mergers detected per year by aLIGO \cite{Postnov&Yungelson14}.   

Population synthesis models [e.g., \cite{Belczynski+08,Postnov&Yungelson14,Dominik+14}] provide independent estimates of the NSNS and NSBH merger rates.  For instance, Dominik et al.~\cite{Dominik+14} predicts NSNS + NSBH detection rates of $\approx 3-7$ yr$^{-1}$ for aLIGO/Virgo, consistent with the empirical estimates above.  However, the full range of rates provided in the literature varies by several orders of magnitude [e.g., \cite{Abadie+10,Postnov&Yungelson14}], due to the large uncertainties in the physics of binary evolution, such as the treatment of common envelope evolution, wind mass-loss from high mass stars, the evolution of metallicity with redshift, and supernova NS kicks.  

NSNS rate calculations usually neglect the influence of external stellar interactions on the evolution of binaries, as justified for the vast majority of stars.  In dense stellar environments, however, such as globular clusters or young stellar clusters, dynamical interactions may enhance the assembly rate of tight NS binaries [e.g.,\cite{Grindlay+06,Samsing+14}].  Additional theoretical uncertainties arise in estimating merger rates in this case due to the poorly constrained evolution of dense stellar systems.  A key aspect of dynamically captured binaries is their potential to merge while the binary orbit still possesses high eccentricity [e.g., \cite{OLeary+09,East+12}].  Although this channel likely represents a small fraction of all mergers \cite{Samsing+14}, even a subdominant population of such events could be of outsized importance to r-process production and kilonova emission ($\S\ref{s:kilonova}$), given the larger ejecta mass from eccentric mergers \cite{East+12}.

\subsection{Precursor Emission}

Compared to the post-merger phase, little study been dedicated to the EM emission during the late inspiral phase prior to coalescence.  If at least one NS is magnetized, then the orbital motion of the conducting companion NS or BH through its dipole magnetic field induces a strong voltage and current along the magnetic field lines connecting the two objects [e.g., \cite{Vietri96,Hansen&Lyutikov01,McWilliams&Levin11,palenzuela2011}].  This voltage accelerates charged particles, potentially powering electromagnetic emission that increases in strength as the orbital velocity increases and the binary separation decreases approaching merger.

Lai \cite{Lai12} shows that the power dissipated by this process is capped by the maximum current which can be sustained before the toroidal magnetic field induced by the current becomes comparable to that of the original poloidal dipole field.  This is because the magnetic pressure of the toroidal field causes the magnetosphere to inflate, tearing open the field lines and disconnecting the circuit.  As an alternative derivation of this maximum power, consider that the Poynting luminosity of a rotating magnetized NS is given by \cite{Contopoulos+99},
\be
L_{\rm P} = \left(\frac{f_{\Phi}}{\tilde{f}_{\Phi}}\right)^{2}\frac{\mu^{2}\Omega^{4}}{c^{3}} \approx 2\times 10^{46}\,\,{\rm erg\,s^{-1}}\,\left(\frac{f_{\Phi}}{\tilde{f}_{\Phi}}\right)^{2}\left(\frac{B_{\rm d}}{10^{13}\,{\rm G}}\right)^{2}\left(\frac{R_{\rm ns}}{12\,{\rm km}}\right)^{6}\left(\frac{P}{\rm ms}\right)^{-4},
\label{eq:LP}
\ee
where $\mu \equiv B_{\rm d}R_{\rm ns}^{3}$, $B_{\rm d}$, $R_{\rm ns}$, and $\Omega = 2\pi/P$ are the dipole moment, surface equatorial dipole field, radius, and rotation rate of the NS, respectively.  Here $f_{\Phi}$ is the fraction of the magnetic flux threading the NS surface which opens to infinity, normalized to its value,
\be 
\tilde{f}_{\Phi} = \int_0^{\theta_{\rm L}}\sin \theta d\theta = [1- \cos\theta_{\rm L}] \approx \theta_{\rm L}^{2} \approx \frac{R_{\rm ns}}{2R_{\rm L}} \approx 0.13\left(\frac{R_{\rm ns}}{12\,{\rm km}}\right)\left(\frac{P}{\rm ms}\right)^{-1},
\ee
for an isolated dipole, where $\theta_{\rm L} \approx \sin^{-1}(R_{\rm ns}/R_{\rm L})^{1/2}$ is the polar latitude of the last closed field line, which intersects the equatorial plane at the light cylinder radius $R_{\rm L} = c/\Omega$.  

Neglecting NS spin, equation (\ref{eq:LP}) can be applied to the case of an orbiting conducting companion by replacing the spin frequency $\Omega$ with the orbital frequency, $\Omega_{\rm orb} = (8GM_{\rm tot}/a^{3})^{1/2}$, where $a$ is the binary separation.  The open fraction $f_{\Phi}$ is reduced by a factor of $\approx 2R_{\rm c}/(2\pi a)$ to account for the azimuthal angle subtended by the binary companion, and increased by a factor of $\approx R_{\rm L}/a = c/(a \Omega_{\rm orb})$ because field lines crossing the equatorial plane exterior to the binary separation $a$ (instead of the light cylinder radius) are now open.  Substituting $f_{\Phi}/\tilde{f}_{\Phi} \approx R_{\rm c}c/(\pi \Omega_{\rm orb} a^{2})$ into equation (\ref{eq:LP}) results in
\be
L_{\rm P} \approx \frac{8\mu^{2}GM_{\rm tot}R_{\rm c}^{2}}{\pi c a^{7}} \approx 3\times 10^{46}\,{\rm erg\,s^{-1}}\,\left(\frac{B_{\rm d}}{10^{13}\,{\rm G}}\right)^{2}\left(\frac{M_{\rm tot}}{2M_{\rm ns}}\right)\left(\frac{R_{\rm c}}{R_{\rm ns}}\right)^{2}\left(\frac{a}{2R_{\rm ns}}\right)^{-7},
\label{eq:Lp}
\ee
where in the numerical estimate we take $M_{\rm ns} = 1.4M_{\odot}$ and $R_{\rm ns} = 12$ km.  Equation (\ref{eq:Lp}) matches that estimated by \cite{Lai12} to within a factor of a few.  The orbital decay time via GW emission is $t_{\rm GW} \equiv a/\dot{a} = 5c^{5}a^{4}/(128G^{3}M_{\rm ns}^{3}) \approx 5(a/2R_{\rm ns})^{4}$ ms, such that the maximum energy released as the orbit decays to separation $a$ is given by
\be
E_{\rm tot} = \int^{a} L_{\rm P}\frac{da}{\dot{a}} \approx \frac{5 \mu^{2}c^{4} R_{\rm c}^{2}}{24\pi  G^{2}M_{\rm ns}^{2}a^{3}} \approx 5\times 10^{43}\,{\rm erg}\left(\frac{B_{\rm d}}{10^{13}\,{\rm G}}\right)^{2}\left(\frac{a}{2 R_{\rm ns}}\right)^{-3}
\ee
This radiation, if thermalized, will likely emerge at hard X-ray/gamma-ray frequencies, possibly resembling a GRB of duration $\sim t_{\rm GW} \sim 1-10$ ms.  However, for magnetic fields of $\lesssim 10^{13}$ G characteristic of radio pulsars, the resulting energy release of $E_{\rm tot} \lesssim 10^{44}$ ergs up to the point of merger ($a = 2 R_{\rm ns}$) is many orders of magnitude smaller than the measured luminosities of short GRBs, events believed to be produced in the merger aftermath ($\S\ref{s:sgrbs}$).

Particles accelerated by magnetospheric interaction could also give rise to a short burst of coherent radio emission \cite{Hansen&Lyutikov01}, perhaps similar to the recently discovered class of events known as ``fast radio bursts" \cite{Thornton+13}.  A low frequency radio precursor could arrive with delay of several seconds or longer after the GWs due to the subluminal group velocity of radio waves through the ionized plasma of the Galactic halo and intergalactic medium.  

In addition to magnetospheric interaction, tidal resonant excitation of modes in the NS crust of provides an additional wave to tap into the orbital energy of the merging binary.  If driven to non-linear amplitudes, such modes shatter the crust, releasing $\approx 10^{46}-10^{47}$ ergs up to tens of seconds prior to merger, potentially producing an observable flare \cite{Tsang+12}.

\section{DYNAMICAL PHASE}
\label{s:dynamical_phase}

The dynamical phase begins when tidal effects become
important, and ends when the central remnant has settled into a stationary 
configuration. The outcome determines the subsequent evolution 
of the remnant and its EM counterparts. Since the changes in the gravitational 
potential over an orbit become non-linear, 
time-dependent simulations are required to answer both qualitative and 
quantitative questions.

Our understanding of this phase has developed through two parallel
approaches: simulations using realistic microphysics with approximate gravity, 
originally aimed at assessing NSNS/NSBH mergers as progenitors 
of short gamma-ray bursts and nucleosynthesis sites, and models that employ full-GR and an 
approximate description of matter, aimed initially at generating
reliable GW predictions. Each approach has contributed
with key insights, and their tracks are gradually converging. 

A number of recent review articles focus on the dynamical aspects
of NSNS/NSBH mergers, primarily from a numerical relativity
perspective \cite{Duez09,shibata2011,Faber&Rasio12,lehner2014}.
Here we provide a brief overview of the key aspects that affect the
prediction of EM counterparts and nucleosynthesis. 

\subsection{Overview of NSNS/NSBH merger evolution}
\label{s:merger_sims_overview}

The final product of the merger is always a central object that contains most 
($\gtrsim 90\%$) of the mass initially in the binary system 
(\S\ref{s:central_remnant_properties}). Matter can be ejected on a 
dynamical time, either by tidal forces and/or by compression at the interface between
objects (\S\ref{s:dynamical_ejecta_formation}), while the remaining
material can possess enough angular momentum to circularize into an accretion 
disk (\S\ref{s:accretion_disk_formation}). 

For quasi-circular NSNS mergers, the key parameter controlling the global evolution
is the binary mass ratio. For nearly equal-mass NSs, the stars deform into symmetric
tear-drop shapes, with relatively small spiral arms, whereas for very
unequal masses the lighter star is tidally disrupted by the 
more massive companion, forming a large spiral arm
[e.g., \cite{bauswein2013}]. For large eccentricities
the dynamics becomes more complex, with one or multiple close encounters before
merger, in which matter can be exchanged and oscillations can be excited due to tidal 
interactions \cite{gold2012}.

For NSBH mergers with small eccentricity, the key parameter is the ratio of the NS
tidal disruption radius to the position of the 
innermost stable circular orbit (ISCO) of the system, 
which depends on the mass ratio, the BH spin 
and the NS radius [e.g., \cite{shibata2011,Foucart12}]. If the tidal radius sits
inside the ISCO radius, the NS is swallowed whole by the BH and no mass
ejection occurs. Otherwise, the neutron star is tidally disrupted,
leading to the formation of long and narrow spiral arms ({\bf Figure~\ref{f:ejecta_geometry}}).
High eccentricity also increases the complexity of the dynamics \cite{Stephens+11}.

Magnetic fields do not appear to have a significant effect on
the global dynamics and gravitational waveform 
during the inspiral and merger [e.g., \cite{giacomazzo2011}]. Instead, they become important
in the subsequent evolution of the merger remnant (\S\ref{s:central_remnant_properties}). 
It has been known for a decade \cite{Price&Rosswog06} 
that strong field amplification can occur at the shearing
interface of NSNS mergers. However, the stringent resolution requirements
demanded by the instabilities involved \cite{obergaulinger2010} are beyond
current computing capabilities for global simulations under
realistic conditions. 

\subsection{Central remnant: prompt versus delayed BH formation}
\label{s:central_remnant_properties}

The maximum mass of a cold, non-rotating neutron star, set by the EOS of 
dense matter, can be increased by rotational and thermal support 
[e.g., \cite{kaplan2014}]. The remnant of a NSNS merger absorbs a significant
fraction of the angular momentum in the binary, and gains thermal energy
via shocks. The non-collapsed, differentially-rotating object is generally 
called a hypermassive neutron star (HMNS). In the case of a NSBH merger, the 
remnant is always a BH.

The fate of the HMNS depends primarily on the total mass of the 
binary \cite{shibata2000}. Above a threshold value, a BH forms on 
a dynamical time ($\sim$~ms). This critical mass is EOS-dependent,
and covers the range $\sim 2.6-3.9M_\odot$ for soft and stiff
nuclear-theory based EOSs, respectively, with empirical relations
connecting the threshold mass with EOS parameters [e.g., \cite{Hotokezaka+11,Bauswein+13}].

If collapse is not prompt, then the lifetime of the HMNS
is determined by a number of processes: 
angular momentum transport via hydrodynamic torques, gravitational
waves [e.g., \cite{shibata2005}] or the magnetorotational 
instability (MRI) [e.g., \cite{Duez+06,Siegel+13}], and loss of thermal
energy via neutrino emission \cite{ruffert1996}. Support against gravity
is eventually lost, with the possible exception of the merger of two low-mass
NSs. The characteristic timescales for angular momentum transport by the MRI 
and neutrino cooling are $\sim 10-100$~ms and $\sim 1-10$~s, 
respectively [e.g., \cite{Paschalidis+12}], while loss of uniform rotational
support for a HMNS slightly above the maximum mass can occur over a much longer 
spin-down timescale. Determining which process is dominant remains an 
active area of research.

The lifetime of the HMNS is key for the qualitative form of the post-merger
GW emission. Prompt BH formation leads to a rapid decrease of the waveform
amplitude (a `ringdown'), while a surviving HMNS generates strong peaks in the GW
spectrum in the frequency range $2-4$~kHz due to sustained global oscillations
[e.g., \cite{sekiguchi2011}]. 
The information contained in these spectral peaks can be used to set constraints on
the EOS of dense matter \cite{bauswein2012a}, although aLIGO-type interferometers 
will only be able to detect them out to few tens of Mpc. An alternative diagnostic of 
the HMNS lifetime is the color of the kilonova (\S\ref{s:kilonova}).

\subsection{Dynamical ejecta}
\label{s:dynamical_ejecta_formation}

The merger gives rise to unbound matter ejection through processes
that operate on the dynamical time, and which depend primarily on the total 
binary mass, the mass ratio, and the EOS. 
Ejected masses lie in the range $10^{-4} - 10^{-2}M_\odot$ for
NSNS mergers [e.g., \cite{Hotokezaka+13}], with velocities $0.1-0.3c$. 
For NSBH mergers, the ejecta can be up to $\sim 0.1M_\odot$, with similar
velocities [e.g., \cite{kyutoku2015}]. More mass can be ejected
if the eccentricity is large \cite{gold2012,East+12}.

\begin{figure}
\centering
 \begin{minipage}{5.5in}
 \begin{minipage}{0.49\textwidth}
     \centering
     \includegraphics[width=\textwidth]{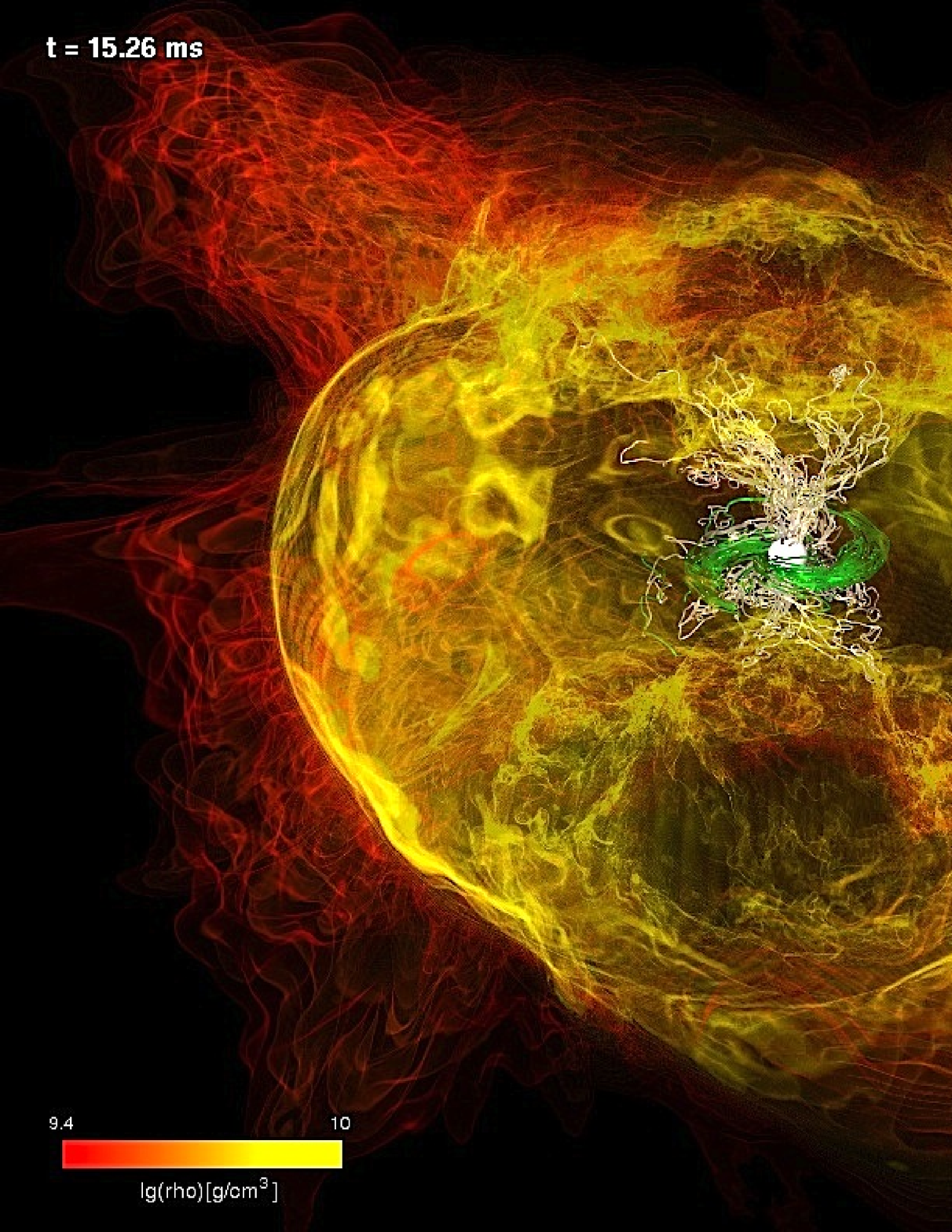}
 \end{minipage}
 \begin{minipage}{0.49\textwidth}
     \centering
     \includegraphics[width=\textwidth]{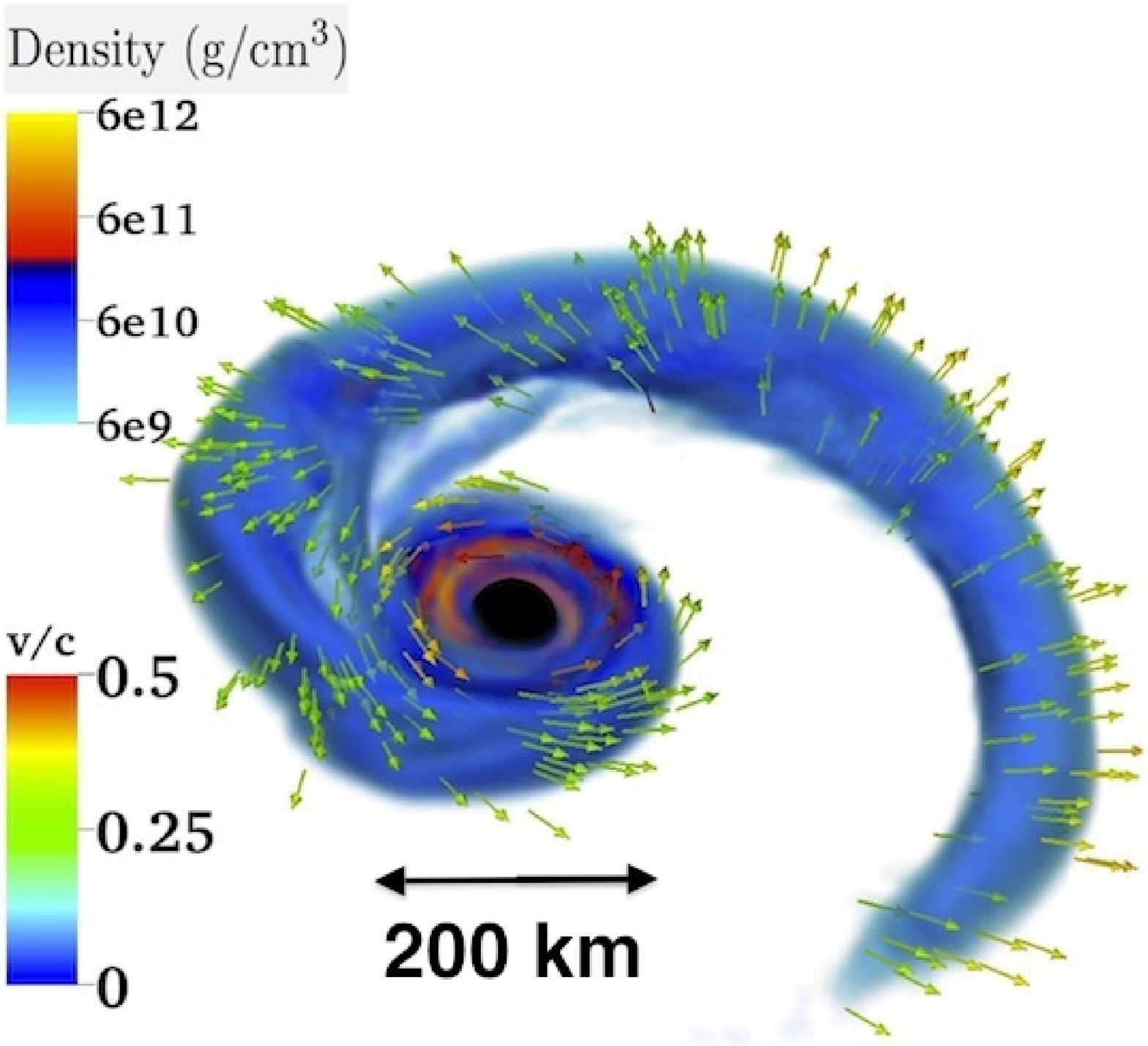}
 \end{minipage}
 \end{minipage}
\caption{Three-dimensional visualizations of the post-merger ejecta.
\emph{Left:} Density field from an equal-mass NSNS merger,
with magnetic field lines shown in green 
[from Rezzolla et al. \cite{Rezzolla+11}, \copyright AAS, reproduced with permission].
The ejecta has significant components in most angular directions (\S\ref{s:dynamical_ejecta_formation}). 
\emph{Right:} Density rendering from a NSBH merger with mass ratio 1.2/7 
[from Foucart et al. \cite{foucart2014}, \copyright Americal Physical Society]. 
The ejecta is confined to the equatorial plane because it is generated primarily
by tidal forces.}
\label{f:ejecta_geometry}
\end{figure}

Two main mass ejection processes operate in NSNS mergers. First,
material at the contact interface is squeezed
out by hydrodynamic forces and is subsequently expelled by 
quasi-radial pulsations of the remnant \cite{oechslin2007,Hotokezaka+13,bauswein2013},
ejecting shock-heated matter in a broad range of angular directions (Figure~\ref{f:ejecta_geometry}). 
The second process involves spiral arms from the non-axisymmetric HMNS, which
expand outwards -- primarily on the equatorial plane -- due to
angular momentum transport by hydrodynamic processes.
The relative importance of these mechanisms depends
on the EOS: a more compact
configuration enhances ejection from
the contact interface \cite{bauswein2013}. The mass ratio
also influences the ejected mass, with very
asymmetric binaries generating up to about twice the
material as a symmetric binary of the same total mass \cite{bauswein2013}.
The total mass also affects the system through the 
BH formation timescale: for prompt collapse, ejection from the contact
interface is suppressed due to prompt swallowing of this
region. Symmetric binaries above the threshold mass yield
some of the lowest dynamical ejecta 
masses ($\sim 10^{-4}M_\odot$) \cite{Hotokezaka+13}.

Regarding NSBH mergers, the primary mechanism of mass ejection
is the tidal force that disrupts the NS on the equatorial
plane via angular momentum redistribution [e.g., \cite{Hotokezaka+13}]. 
The geometry of the ejecta is thus fundamentally different from that of NSNS mergers,
as Figure~\ref{f:ejecta_geometry} illustrates. 
Also, the ejecta from NSBH mergers often covers only part of the azimuthal range \cite{kyutoku2015}. 
This has important implications for the properties of the
radioactively-powered electromagnetic counterparts (\S\ref{s:kilonova}).

The thermodynamic properties of the dynamical ejecta can depend
sensitively on the microphysics employed and the treatment of
neutrinos, particularly for NSNS mergers. Studies in full
GR with all these components have only recently been performed,
remaining an active area of research
[e.g., \cite{wanajo2014,foucart2015a}].

\subsection{Accretion disk formation}
\label{s:accretion_disk_formation}

Ejected material that is not gravitationally unbound from 
the central remnant either falls back or circularizes
into an accretion disk. The disk mass depends primarily on the 
mass ratio of the binary, and on the spins of the binary 
components, the EOS, and the total binary mass [e.g., \cite{oechslin2007,shibata2011}].
For quasi-circular NSNS and NSBH mergers, disk masses can 
reach up to $\sim 0.3M_\odot$ [e.g., \cite{oechslin2007,Hotokezaka+13c,foucart2014}],
with a similar range obtained for large eccentricities \cite{Stephens+11,gold2012}.

In NSNS mergers, the material that forms the disk comes from 
the contact interface and the tips of the spiral arms \cite{Oechslin&Janka06}. The
disk mass appears to be particularly
sensitive to the development of a primary spiral
arm, with a corresponding sensitivity to the mass ratio 
\cite{oechslin2007}. The spins of the NS enter through 
the total angular momentum in the pre-merger binary: 
co-rotating spins yield larger disk masses
than irrotational binaries \cite{oechslin2007}. A stiffer
EOS (larger NS radius for fixed mass) leads to 
more efficient tidal disruption for asymmetric binaries, 
and hence to larger spiral arms and thus higher disk mass \cite{oechslin2007}.
The total binary mass influences the disk through the BH formation
time: prompt collapse quickly swallows mass from regions that
would otherwise end up as disk material \cite{ruffert1999}.
The lowest disk masses ($\sim 10^{-4}M_\odot$)
are obtained for symmetric binaries above the
threshold for prompt BH formation \cite{Hotokezaka+13}.
In contrast, a longer-lived HMNS can transfer
angular momentum outwards via dynamical torques
for a sustained period, leading to a larger 
disk mass [e.g., \cite{Hotokezaka+13c}].

Regarding NSBH mergers, the disk forms when the tidally-stretched 
neutron star wraps around and intersects itself 
[e.g., \cite{shibata2011}], with the possible exception of systems with
precession. Despite the presence of
shocks, a one-armed spiral structure persists for many
orbits in the disk, transporting angular momentum and generating
accretion onto the BH [e.g., \cite{kyutoku2015}].
In addition to the dependence on the stiffness
of the EOS as in NSNS mergers, the disk mass also
increases for larger spin in the initial BH. 

\section{ACCRETION PHASE}
\label{s:accretion}

Once a few dynamical times have elapsed, the gravity of the central remnant settles 
into a stationary form and the subsequent evolution of the system occurs on longer
timescales. 

\subsection{Short Gamma-Ray Burst Connection}
\label{s:sgrbs}

\subsubsection{Observational Connection}

Gamma-ray bursts (GRBs) are luminous and highly variable flashes of $\sim$ MeV
$\gamma$-rays.  Long-duration GRBs, usually defined as those lasting longer
than 2 seconds, are associated with the core-collapse of massive stars
[e.g., \cite{Woosley&Bloom06}].
Short-duration GRBs (SGRBs) last less than two seconds and are
characterized by spectrally harder emission.  Their origin is poorly understood
compared to long bursts, due in part to their dimmer afterglows, which makes
them more challenging to localize.

GRBs are produced by internal dissipation and non-thermal emission within a
collimated ultra-relativistic outflow, or `jet'.  Their rapid variability and
enormous isotropic-equivalent luminosities of $L_{\gamma} \sim 10^{50}-10^{52}$ erg
s$^{-1}$ demand that GRB jets be powered by rotational or gravitational energy
released from a solar mass compact object.  This energy release could occur
through the accretion of a massive torus
on the characteristic accretion (``viscous") timescale
\be
t_{\rm visc} \sim \frac{R^{2}}{\nu} \approx 0.26\,{\rm s}\,\left(\frac{\alpha}{0.1}\right)^{-1}\left(\frac{R}{\rm 30\,km}\right)^{3/2}\left(\frac{M_{\rm BH}}{3M_{\odot}}\right)^{1/2}\left(\frac{H}{0.1R}\right)^{-2},
\label{eq:tvisc}
\ee
where $R$ is the characteristic radial extent of the torus, $\nu = \alpha
c_{\rm s}H$ is the effective viscosity due to 
turbulence within the disk,
$c_{\rm s}$ is the midplane sound speed, 
and $H \approx c_{\rm s}/\Omega_{\rm K}$ is the vertical
scale-height of the disk,
with $\Omega_{\rm K} = (R^{3}/GM_{\rm BH})^{1/2}$ 
the Keplerian orbital frequency around a BH of mass $M_{\rm BH}$.

For NSNS/NSBH mergers, we expect $M_{\rm BH} \sim 3-20M_{\odot}$, $R \sim$
tens of km, such that $t_{\rm visc} \sim 0.1-1$ s for a range of physical
viscosities $\alpha \sim 0.01-0.1$.  The correspondence between $t_{\rm acc}$
and the observed durations of SGRBs provides suggestive evidence for a link
between SGRBs and NSNS/NSBH mergers \cite{Paczynski86, Eichler+89}.  Additional
observational support of the SGRB-merger connection 
includes [e.g., \cite{Berger14}]: (1) an association with both
spiral and elliptical host galaxies, 
consistent with the expected distribution of delay times between star formation and NSNS/NSBH mergers; 
(2) a lack of bright coincident supernovae or other evidence of star formation,
indicating a non-massive star origin;
(3) a distribution of the observed spatial offsets of SGRBs from their host
galaxy light which is consistent with that resulting from NS birth kicks
\cite{Belczynski+06,Fong&Berger13}; and
(4) the possible detection of kilonova emission 
following the SGRB 130603B [\cite{Tanvir+13,Berger+13}, $\S\ref{s:kilonova}$].

Equation (\ref{eq:tvisc}) shows that producing a SGRB
through accretion requires 
forming a massive, compact torus
on a comparable timescale,
disfavoring scenarios that would produce a more
radially-extended disk, such as
the collapse of a massive star.  One model which can 
overcome this limitation is the accretion-induced collapse of a NS [e.g.,
\cite{Dermer&Atoyan06}].  However, 
such an event is unlikely to leave a remnant disk outside of the BH horizon for NS properties
consistent with observationally allowed EOSs \cite{Margalit+15}.

\subsubsection{Relativistic Jet Formation}

The SGRB jet energy can be parameterized as $E_{\rm j} = \epsilon_{\rm j}M_{\rm
t}c^{2} \approx 10^{50}(\epsilon_{\rm j}/10^{-3})(M_{\rm t}/0.1M_{\odot})$ erg,
where $M_{\rm t}$ is the torus mass and $\epsilon_{\rm j}$ is an efficiency
factor.  For a torus mass
$M_{t} \sim 0.1M_{\odot}$
($\S\ref{s:accretion_disk_formation}$),
producing a GRB jet of the observed energy range of $E_{\gamma} \approx
L_{\gamma}t_{\gamma}f_{\rm b} \sim 10^{48}-10^{51}$ erg requires only a modest
efficiency factor $\epsilon_{\rm j} \lesssim 10^{-5}-10^{-2}$, where $f_{\rm b}
\sim 0.01-0.1$ is the uncertain beaming fraction that relates the true energy
to that assuming isotropic emission based on the observed flux,
and $t_\gamma$ is the burst duration.  In order to
explain the high inferred jet Lorentz factors of $\Gamma \gtrsim 100$, this
energy must furthermore be placed into a small quantity of baryonic mass
$M_{\rm j} \approx E_{\rm j}/\Gamma c^{2} \approx 10^{-7}-10^{-5}M_{\odot}$.  

The relatively low density of the polar region above the 
central remnant
provides a natural location for jet
launching.  Two energy sources are usually considered.
First, neutrinos from the accretion
torus deposit a fraction of their energy along the polar axis via self
annihilation, mainly $\nu_{e}-\bar{\nu}_{e}$ \cite{Goodman86}.  This `high
entropy' jet is accelerated to relativistic velocities by thermal pressure, as
in classical fireball models \cite{Paczynski86}.  The strength of this model is
that annihilation heating must occur at some level.
A weakness, however, is
the low efficiency $\epsilon_{\rm j} \lesssim 10^{-3}$, 
which
furthermore peaks at the very earliest times after the merger, 
when the dense dynamical and wind-driven ejecta could inhibit jet formation
\cite{Murguia-Berthier+14,Just+15b}.

The jet could alternatively be powered by the Poynting flux from a strong
magnetic field threading the rotating central compact object or the accretion
disk.   An MHD jet can in principle possess a much higher efficiency
($\epsilon_{\rm j} \sim 1$), especially if the power source is the spin energy
of the black hole \cite{Blandford&Znajek77}.  
Dynamical simulations of NSNS/NSBH mergers have begun to
explore magnetic field amplification following the merger
[e.g., \cite{Kiuchi+14}, see also \S\ref{s:merger_sims_overview}]
with some showing the formation
of a magnetic field topology conducive to jet formation
\cite{Rezzolla+11,Paschalidis+15}.  This represents an important direction for
future research.

After being launched near the central compact object, the jet may be collimated
on larger scales by confinement provided by a dense external medium.  This
medium may be provided by the dynamical ejecta or accretion disk outflows
\cite{Aloy+05, Bucciantini+12,Nagakura+14}.

\subsubsection{Extended Emission and Millisecond Magnetar Remnants}
\label{s:EE}

Roughly one quarter of SGRBs are accompanied by temporally extended hard X-ray
emission, which lasts for a minute or longer after the main burst
\cite{Norris&Bonnell06}.  The rapid variability of this ``extended emission"
(EE) indicates that it is powered by on-going activity from the central engine,
as opposed to the GRB afterglow at much larger radii.  The total isotropic
energy of the EE is comparable or exceeding that of the initial SGRB itself.
Some SGRBs are also accompanied by X-ray afterglows that decay more slowly in
time than  predicted by synchrotron blast wave models \cite{Rowlinson+13}. 
The abrupt cut-off of this plateau-like emission in some cases is again
challenging to explain with the afterglow, implying a central engine origin
\cite{Rowlinson+13}.

The EE and X-ray plateaus last much longer than the expected lifetime of the
prompt accretion disk (eq.~[\ref{eq:tvisc}]), posing a challenge to standard
NSNS/NSBH merger models.  This activity could be powered by the accretion of
marginally bound matter which falls back to the BH at late times [e.g.,
\cite{Rosswog07,Lee+09}].  However, fall-back models 
require very low disk viscosities
\cite{Lee+09} 
or a magnetic field topology that plays
a non-trivial role in the jet dynamics \cite{Kisaka&Ioka15}.
Accretion disk outflows could 
also obstruct the fall-back material \cite{Fernandez+15}.   

An alternative model postulates that the EE is powered by a rotating magnetized
NS remnant, a `millisecond magnetar'.  The discovery of $\sim 2M_{\odot}$ NSs
[e.g., \cite{Antoniadis+13}] makes it likely that NSNS mergers produce
meta-stable HMNS remnants ($\S\ref{s:central_remnant_properties}$), if not 
NSs which are indefinitely stable to gravitational collapse
[e.g.~\cite{Ozel+10,Bucciantini+12,Giacomazzo&Perna13}].  The NS remnant is
formed with a rotational period $P \sim 1$ ms 
and may acquire a strong magnetic field $\gtrsim 10^{14}-10^{15}$ G via 
dynamo action \cite{Duncan&Thompson92}.  The NSNS merger
remnant possesses a reservoir of rotational energy of $E_{\rm rot} =
I\Omega^{2}/2 \simeq 3\times 10^{52}{\rm erg}\left(P/1\,\rm ms\right)^{-2},$
where $I \simeq 2\times 10^{45}$ g cm$^{2}$ is the NS moment of inertia, which
can be extracted 
via magnetic dipole radiation (eq.~[\ref{eq:LP}]).
If the remnant is initially supported by rotation, then once enough angular
momentum is lost to EM emission or GWs, its resulting collapse to a BH could
result in an abrupt decrease in the X-ray emission \cite{Rowlinson+13}.  

A transient
relativistic jet could be powered by the remnant torus accreting onto the NS
remnant (instead of a BH), in analogy with Galactic accreting NS systems with
relativistic jets [e.g.~Circinus X-1].  This scenario would naturally preserve
the correspondence between the viscous timescale (eq.~[\ref{eq:tvisc}]) and the
GRB duration.  However, baryon contamination by the neutrino-driven
wind (\S\ref{s:neutrino_driven_wind}) could prevent the jet
from becoming ultra-relativistic [e.g., \cite{Dessart+09}].  
If most of the field lines threading the rotating NS pass through the
accretion disk or a slower wider angle wind instead of the core of the jet,
then the jet baryon loading would be correspondingly reduced. 
Until the interplay between neutrinos and the
magnetic field is better understood in this complex, highly time-dependent
environment, it may be premature to conclude that BH formation is necessary to
produce a SGRB.  

\subsection{Late Disk Evolution and Outflows}

\subsubsection{Global evolution}
\label{s:accretion_disk_timescales}

The basic properties of neutrino-cooled accretion disks
were established by \cite{popham1999} using steady-state
models. For the relevant accretion rates, 
there is a near balance between heating by angular
momentum transport and neutrino losses at small radii.
At larger radii, where the temperature is lower, neutrino 
cooling becomes inefficient and the disk reaches the `advective' 
regime \cite{Narayan&Yi94}. 
The transition occurs approximately where the viscous time [equation~(\ref{eq:tvisc})]
is equal to
the neutrino cooling time ($T\sim 3$~MeV for $\alpha = 0.03$
and typical disk parameters). 
For high accretion rates ($\gtrsim 1M_\odot$~s$^{-1}$),
the disk becomes opaque to neutrinos and it also cools
inefficiently [e.g., \cite{Chen&Beloborodov07}].

The first time-dependent calculation that followed the disk evolution over
several viscous times using a one-zone model \cite{metzger2008steady}
showed that as the disk evolves, it transitions from the neutrino-cooled stage 
to the advective state as it spreads and its temperature decreases. 
These results were confirmed by subsequent one-dimensional height-integrated 
models \cite{Metzger+09a}.

Multi-dimensional simulations were
first carried out by \cite{ruffert1999} for a few orbits ($\sim 10$~ms), solving 
the hydrodynamic equations in three-dimensions, with a physical EOS and 
a neutrino leakage scheme. Results showed that the basic expectations
about neutrino opaqueness and cooling efficiency from
\cite{popham1999} were correct. Work by Lee et al.
\cite{lee2002,lee2004}
included for the first time
angular momentum transport by an $\alpha$ viscosity, and carried out
the evolution for a few viscous times ($\sim 1$~s), showing
that qualitative transitions in the evolution can occur depending on the
relation between the viscous time and the neutrino cooling time.
The three dimensional models of \cite{setiawan2006} also 
evolved the disk to a time $\sim 400$~ms and characterized the effect 
of viscosity and BH spin on the neutrino emission. Finally, the
disk evolution in MHD has been computed to times $\sim 100$~ms 
by \cite{shibata2007,shibata2012}.

\begin{textbox}

\subsubsection{Neutrino Signal} 
\label{s:neutrino_signal}

Neutrino emission from NSNS/NSBH mergers is comparable
to that from core-collapse supernovae, with peak luminosities 
$\sim 10^{53}$~erg~s$^{-1}$ \cite{ruffert1997}. The duration of this signal can vary. 
If a BH forms promptly, the emission is dominated by accretion and
hence it decreases on a viscous time (equation~\ref{eq:tvisc}), while a long-lived
HMNS significantly increases the magnitude and duration of the signal.

The most optimistic rate of NSNS mergers from \cite{Abadie+10} yields 
one event per $1000$~yr in the galaxy, or a factor at least $\sim 10$ lower than the 
core-collapse supernova rate. Current observatories
can at best detect $\sim 1$ neutrino from a supernova in
Andromeda (distance $780$~kpc)  \cite{scholberg2012}. Hence it is unlikely that neutrino
emission will be observed from a NSNS/NSBH merger before 
megaton-scale detectors become available.
\end{textbox}

\subsubsection{Mass ejection by neutrino energy deposition} 
\label{s:neutrino_driven_wind}

Duncan et al. \cite{duncan1986} showed that when the neutrino
temperature of a protoneutron star exceeds a threshold, a hydrostatic
atmosphere is no longer possible and a thermally-driven wind is launched.
Ruffert et al. \cite{ruffert1997} pointed out that
this phenomenon could also occur in the remnants of NSNS/NSBH mergers,
with neutron-rich material being ejected in this `neutrino-driven wind'.
A series of papers by Surman, McLaughlin, and collaborators 
[e.g., \cite{Surman+08}]
used parameterized wind trajectories to explore 
$r$-process production in these outflows, with similar
work carried out by \cite{Metzger+08a,Wanajo&Janka12}.

Time dependent models of remnant tori that include neutrino
emission and absorption indicate that when a BH forms
promptly, the amount of mass ejected through this channel 
is very small because the torus becomes transparent 
very quickly \cite{setiawan2006}. 
Fern\'andez \& Metzger \cite{Fernandez&Metzger13} evolved disks around 
BH for a time $\sim 1$~s,
finding that neutrino energy deposition is dynamically
unimportant compared to viscous energy 
deposition (\S\ref{s:viscously_driven_wind}).
This result was confirmed by \cite{Just+15} using more 
advanced neutrino transport, with an estimated
contribution to the total
disk outflows of $\sim$ few \% due to neutrino heating alone.

\begin{figure}
\includegraphics*[width=5.5in]{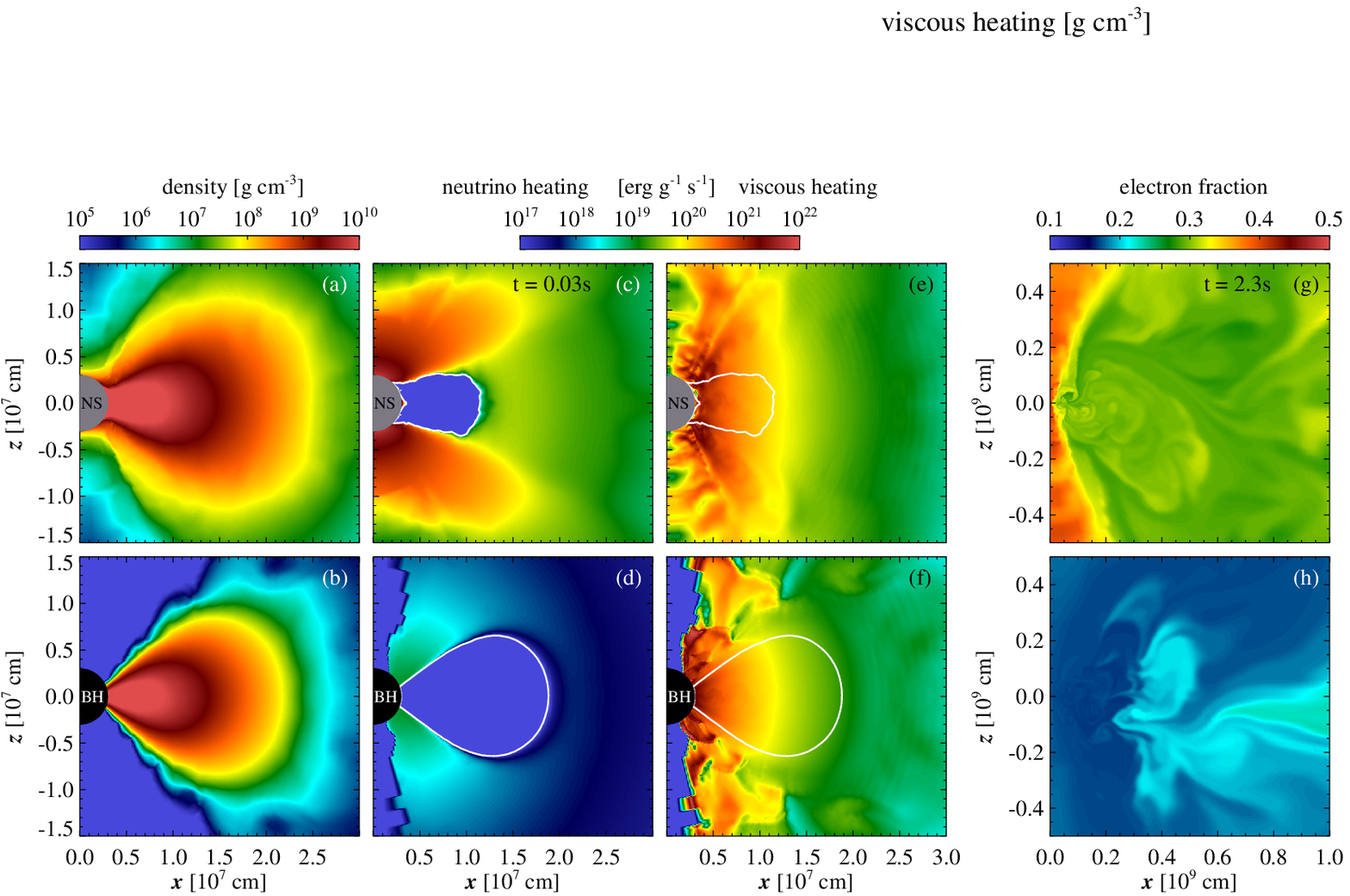}
\caption{Snapshots in the evolution of NSNS/NSBH remnant accretion disks
around a HMNS (top row) and BH (bottom row), highlighting the differences
in key quantities.
Columns show density, neutrino heating, and viscous heating at a time
comparable to the thermal time in the disk ($\sim 30$~ms), and
the electron fraction at later times when the disk wind is underway (note
the difference scale of the rightmost column). Figure from Metzger \& Fern\'andez \cite{Metzger&Fernandez14}.}
\label{f:hmns_bh_comparison_M14}
\end{figure}

If the HMNS can survive for
longer than a thermal time in the inner disk ($\sim 50$~ms),
the larger neutrino luminosity [{\bf Figure~\ref{f:hmns_bh_comparison_M14}}, see
also \cite{richers2015}]
can lead to the ejection of non-negligible amounts
of mass ($\sim 10^{-3}M_\odot$). This was first shown explicitly by \cite{Dessart+09}
using two-dimensional simulations with neutrino
transport, and has recently been revisited by \cite{perego2014} 
using three-dimensional simulations. 

\subsubsection{Late mass ejection in the advective state}
\label{s:viscously_driven_wind}
\begin{figure}

\includegraphics*[width=0.7\textwidth]{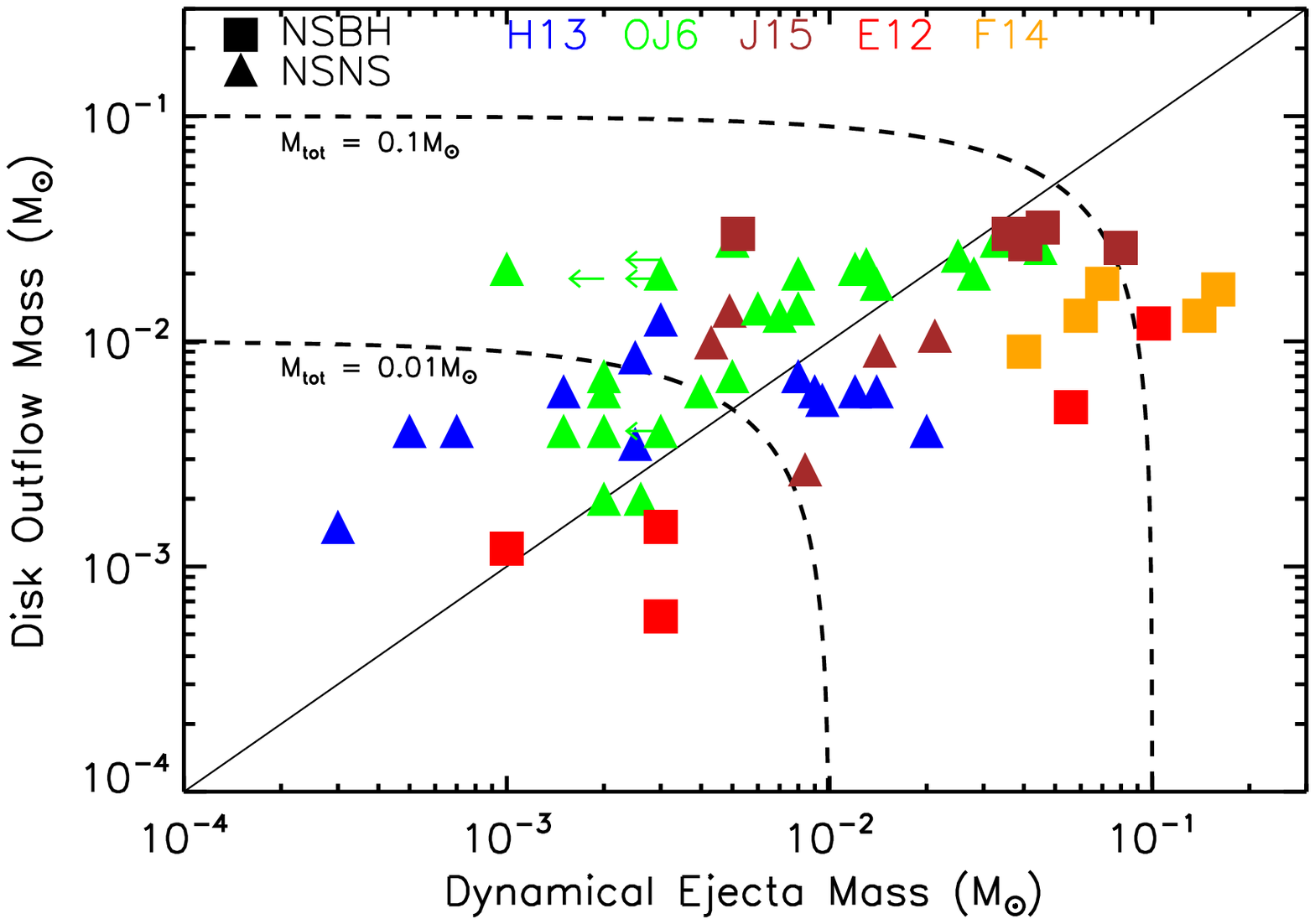}
\caption{Mass ejected dynamically versus that ejected 
in disk outflows.  Each point corresponds to the results of a single
simulation, including the NSNS models of 
\cite{Hotokezaka+13} ({\it blue triangle}), 
\cite{Oechslin&Janka06} ({\it green triangles}; upper limits are shown as arrows), 
\cite{Just+15} ({\it brown triangles}) and the
NSBH models of \cite{East+12} ({\it red squares}) and 
\cite{foucart2014}. The mass 
in disk outflows is estimated to be 10 per cent of the disk mass (\S\ref{s:viscously_driven_wind}).
Dashed lines show a combined ejecta mass (dynamical + disk winds) of $\approx 0.1 M_{\odot}$
and $0.01M_{\odot}$, 
showing the range necessary to explain the Galactic
production rate of heavy $r$-process nuclei $\sim 5\times 10^{-7}M_{\odot}$
yr$^{-1}$, given the allowed range of rates of NSNS mergers $\in [4,61]$
Myr$^{-1}$ (99$\%$ confidence) as calculated based on the population of
Galactic binaries \cite{Kim+15}.}
\label{f:disk_dynamical_mass_metzger}
\end{figure}

Significant outflows can be driven on timescales longer
than the thermal time once the bulk of the disk achieves the 
advective state, in which heating by angular
momentum transport and nuclear recombination are unbalanced by neutrino cooling 
[e.g., \cite{kohri2005}].
This state is accompanied by freezout of weak interactions,
leading to a neutron-rich wind \cite{metzger2008steady,Metzger+09a}.

The work of \cite{Fernandez&Metzger13} addressed
the late-time evolution of the disk
and its composition simultaneously, accounting for neutrino
emission and absorption. Results showed that for a non-spinning BH, 
several percent of the initial disk mass are ejected on 
a timescale of $\sim 1$~s, and that the material is sufficiently neutron-rich
($Y_e \sim 0.2$) to produce heavy $r$-process elements.
Just et al. \cite{Just+15} included two-moment 
neutrino transport and the effects of BH spin, finding more mass ejection (factor of several)
when the BH spin has a more representative value $\chi \simeq 0.8$, 
with electron fractions as high as $Y_e \sim 0.35$. 

A larger amount of mass ($\gtrsim 90\%$ of the initial disk mass) 
can be ejected when a long-lived HMNS sits at the center \cite{Metzger&Fernandez14}.
The difference is related to the presence of a hard surface and the
higher level of neutrino irradiation (Figure~\ref{f:hmns_bh_comparison_M14}).
This also leads to a higher average $Y_e$ in the wind and thus
to a variable composition that depends on the lifetime of the 
HMNS.

Currently, the largest theoretical uncertainty is
the magnitude of the heating due to angular momentum transport.
The work of \cite{shibata2007,shibata2012} has pioneered
the use of MHD in this context, but the evolution
was not long enough and the floor of density too high for
the appearance of the late-time outflow.

The mass ejected by the late disk
wind can be comparable to or larger than that in the dynamical ejecta.
This is illustrated in {\bf Figure~\ref{f:disk_dynamical_mass_metzger}}, 
where dynamical ejecta masses from several merger simulations are compared with estimates
for the disk wind, assuming $10\%$ of the initial disk mass.

\section{$r$-PROCESS NUCLEOSYNTHESIS AND ITS SIGNATURE (KILONOVA)}
\label{s:r-process}

In this section we briefly summarize the basic properties of
$r$-process nucleosynthesis in NSNS and NSBH mergers, and the
associated EM counterparts.
Recent reviews on $r$-process nucleosynthesis
can be found in \cite{arnould2007,sneden2008,mumpower2015}, 
and on kilonovae and related EM emission in \cite{Metzger&Berger12,Rosswog15}. 
The reader should also consult the article 
by Arcones \& Mart\'inez-Pinedo  in this volume.

\subsection{$r$-Process Nucleosynthesis in NS Mergers}
\label{s:nucleosynthesis_overview} 
 
The rapid neutron capture process (`$r$-process') was
introduced to account for elements 
which require formation through neutron captures 
at rates faster than beta decays, given the properties of the nuclear chart and
the Solar System abundance curve
\cite{BBFH57,cameron1957}. Approximately half of the elements
heavier than Zinc ($Z=30$) are thought to be generated 
through this channel [e.g., \cite{sneden2008}].

The large neutron densities, high temperatures, and fast timescales
required suggest an explosive
environment. Core-collapse supernovae have been a 
leading candidate since the earliest $r$-process studies, with the neutrino
driven wind phase becoming the focus as understanding of the explosion 
improved \cite{meyer1992}. 
Additional candidate sites include the jets of 
magnetorotational supernovae \cite{cameron2001}
and a neutrino-induced $r$-process in the He shells of massive stars at late 
times during core-collapse supernovae \cite{epstein1988}.

NSBH mergers were first suggested as a site for the 
$r$-process by Lattimer \& Schramm 
\cite{Lattimer&Schramm74} by virtue of
material ejected by the tidal disruption of the NS.
They noted that the total yield from these events
could match the amount of $r$-process elements observed in the Galaxy. 
Lack of knowledge about the dense matter EOS, the approximate
calculation method, and uncertain statistics about the
merger rates prevented more definitive statements.
Significant progress had to wait more than 20 years: 
Freiburghaus et al. \cite{Freiburghaus+99} ran a nuclear reaction 
network on tracer particles from Newtonian
NSNS merger simulations with a physical EOS, finding 
good agreement with the solar system $r$-process abundances
for suitable initial $Y_e$ of the material. 

\subsubsection{Nucleosynthetic Yields }
  \label{s:r-process_yields} 

The dynamical ejecta from NSNS/NSBH mergers can be a robust generator 
of heavy ($A\gtrsim 140$) $r$-process elements.
This robustness is rooted in
`fission recycling' [e.g., \cite{goriely2005}]:
the low initial $Y_e$ results in a large
neutron-to-seed ratio, allowing the nuclear flow to reach
heavy nuclei for which fission is possible ($A\gtrsim 250$).
The fission fragments are then subject to additional neutron
captures, generating more heavy nuclei and closing the cycle.

\begin{figure*}
\centering
 \begin{minipage}{5.5in}
 \begin{minipage}{0.49\textwidth}
     \centering
     \includegraphics[width=\textwidth]{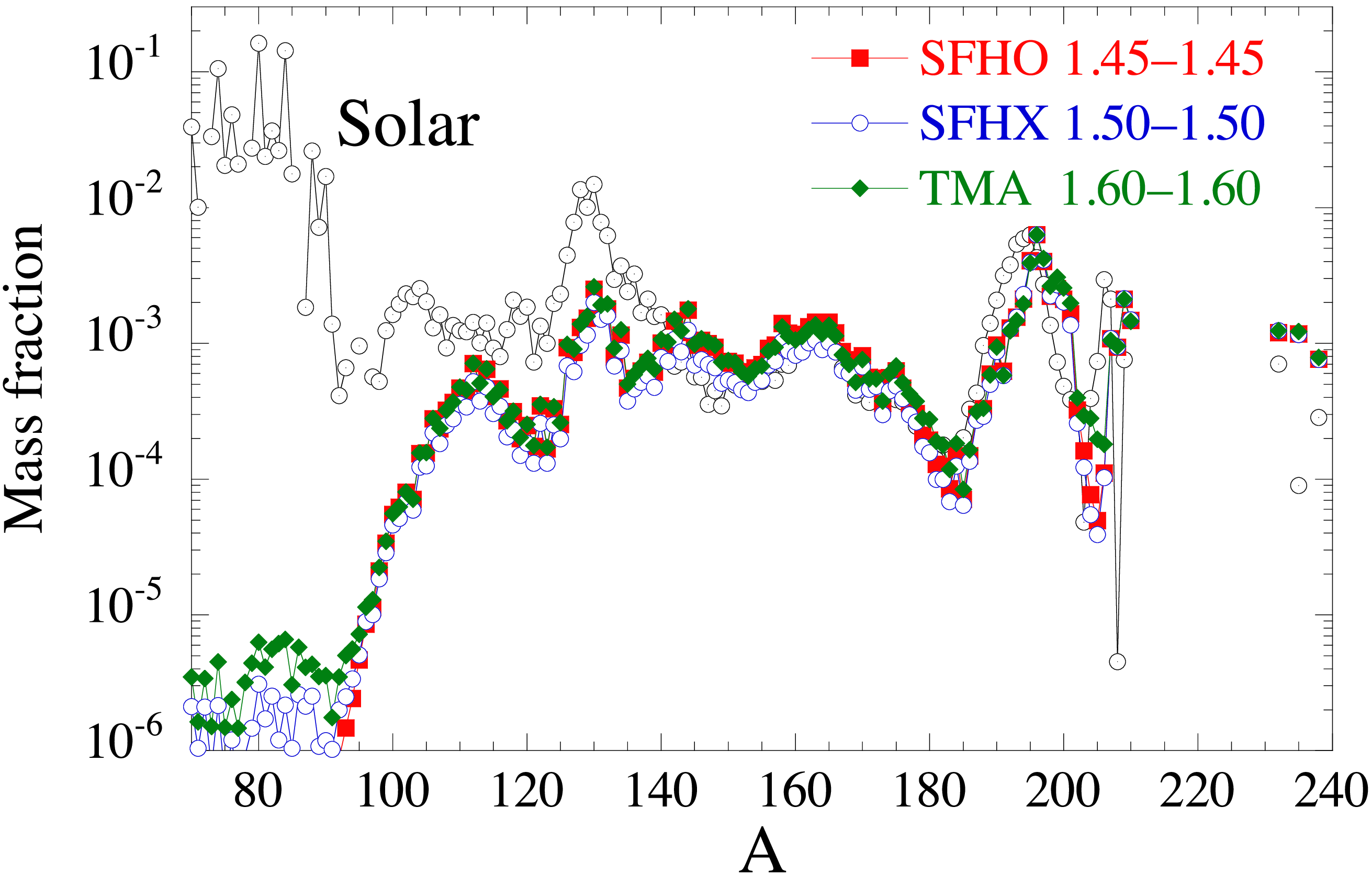}
 \end{minipage}
 \begin{minipage}{0.49\textwidth}
     \centering
     \includegraphics[width=\textwidth]{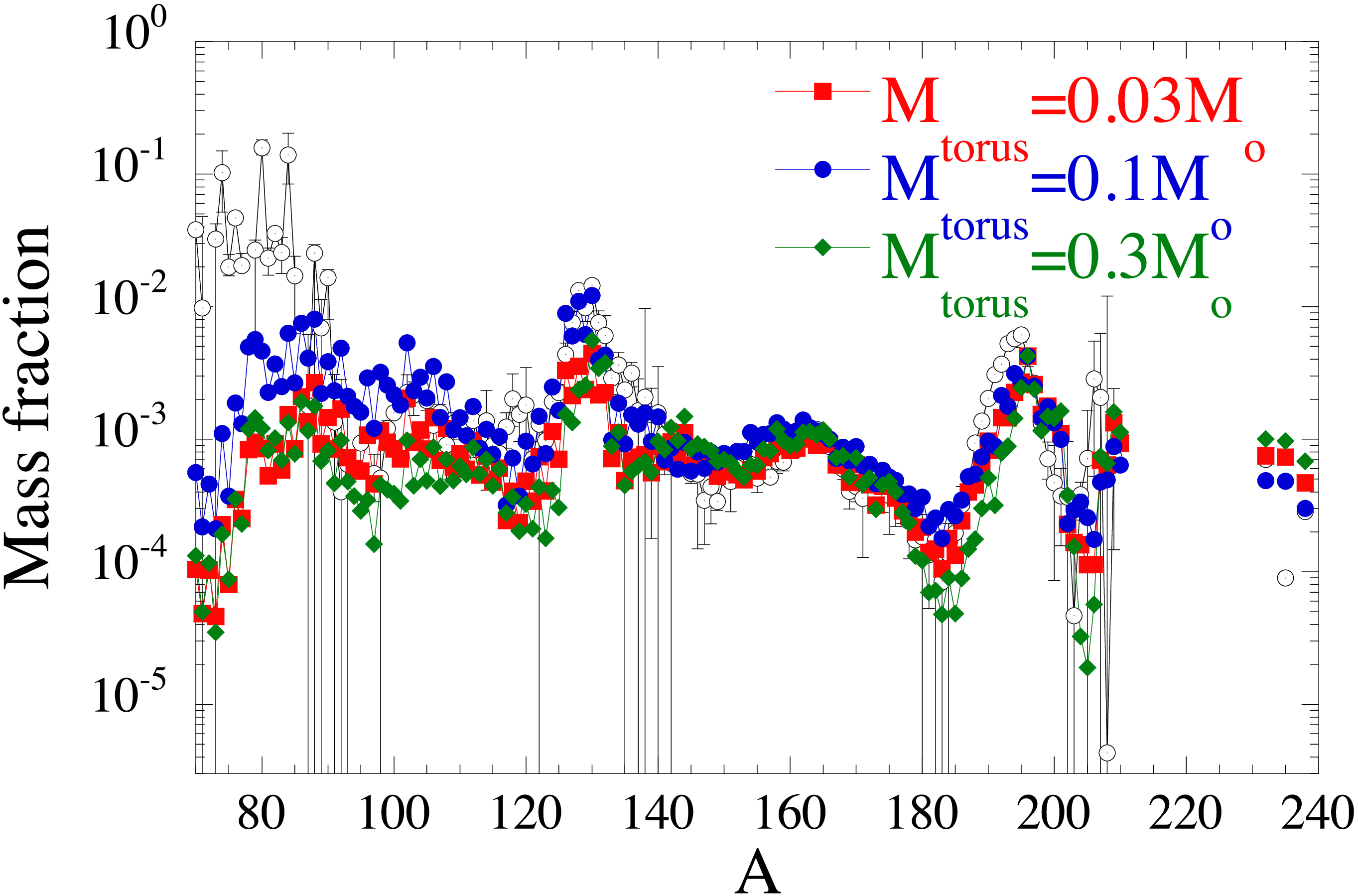}
 \end{minipage}
 \end{minipage}
\caption{\emph{Left:} Mass fractions synthesized during the expansion of the
dynamical ejecta for three equal-mass, delayed-collapse NSNS merger models that 
employ different equations of state and total binary masses. Circles denote
the Solar System $r$-process mass fractions. Note the low sensitivity of the
abundances to binary parameters or EOS. \emph{Right:} Combined mass
fractions produced by the dynamical ejecta and disk wind for three disk
masses, showing the dispersion in the abundance of elements with $A\lesssim 140$
due to the different disk masses. Figures from Just et al. \cite{Just+15}, reproduced
by permission of Oxford University Press on behalf of the Royal Astronomical Society}
\label{f:just_combined_abundances}
\end{figure*}

The resulting abundance pattern follows the general shape
of the $r$-process distribution in the Solar System and metal-poor halo stars, 
with abundances below $A\sim 140$ depleted \cite{Freiburghaus+99}. The quantitative
agreement depends primarily on nuclear physics properties such as
the nuclear mass model, fission fragment distribution, and $\beta$-decay
half lives [e.g., \cite{eichler2015}].
Other factors such as the EOS or binary
parameters, while key for determining the amount of mass ejected (\S\ref{s:dynamical_ejecta_formation}),
have little influence on the abundance distribution \cite{Goriely+11,Korobkin+12,bauswein2013},
as illustrated in {\bf Figure~\ref{f:just_combined_abundances}}.

Elements with $A\lesssim 140$ can also
be generated in NSNS/NSBH mergers, although the yields are sensitive to
astrophysical parameters. The 
late-time wind
can synthesize elements in the entire $r$-process range due to 
its broad $Y_e$ distribution \cite{Just+15}. When
combined with the dynamical ejecta,
the ensemble distribution contains an intrinsic 
dispersion for $A\lesssim 140$, with a dependence on parameters such as the
disk mass (Figure~\ref{f:just_combined_abundances}). In the case
of a long-lived HMNS, the neutrino-driven wind is also
a source of light $r$-process elements \cite{martin2015}. Finally, recent work 
that includes neutrino absorption in dynamical merger simulations  
finds that irradiation from the HMNS 
can generate a broad $Y_e$ distribution, leading to 
the entire range of $r$-process elements being generated
by the dynamical ejecta alone \cite{wanajo2014}.

\subsubsection{Observational Constraints }
\label{s:r-process_constraints} 

By now it is well-established observationally that many old, metal-poor stars 
in the Galactic halo have inferred $r$-process element abundance distributions 
that closely follow the Solar System pattern for elements heavier than Ba ($Z\geq 56$), 
particularly at low metallicity [e.g., \cite{sneden2008}]. This is taken as evidence
for a robust $r$-process site that operated since early on in the life of the Galaxy.

One standard observational test to be met by any candidate site is
the enrichment history of the Galaxy, 
as inferred from the Eu/Fe abundance ratio (relative to solar) 
as a function of metallicity in metal-poor halo stars.
Recent studies that include realistic mixing 
processes in galaxy simulations \cite{vdvoort2015,shen2015,hirai2015} find reasonable 
agreement with the observations when assuming that NSNS/NSBH mergers are the 
dominant source of $r$-process enrichment.

Regarding light $r$-process elements, the abundance patterns inferred
from metal-poor stars do not uniformly follow the Solar System, with a
dispersion in the relative amounts of heavy and light elements [e.g., \cite{sneden2008}].
NSNS/NSBH mergers are able to provide for light $r$-process elements through
disk outflows, in amounts that have an intrinsic dispersion relative to heavier elements
due to the sensitivity of the disk and dynamical ejecta masses to binary parameters \cite{Just+15}.
Current models of CCSNe nucleosynthesis also yield primarily light $r$-process elements
in the neutrino-driven wind, contributing further to the diversity of 
light $r$-process element abundances in Galactic stars [e.g., \cite{arcones2011}].


\subsection{Kilonova}
\label{s:kilonova}

Nuclei freshly synthesized by the r-process are radioactive.  As this expanding
matter decays back to 
stability,
the energy released via beta-decays
and fission can power a thermal transient lasting days to weeks,
commonly\footnote{The term `kilonova' was introduced by Metzger et al.~\cite{Metzger+10},
who first calculated light curves self-consistently using the radioactive
heating of r-process nuclei.  The term `macronova' is also sometimes used
\cite{Kulkarni05}.} known as a `kilonova'
\cite{Li&Pacyznski98,Kulkarni05,Metzger+10,Roberts+11}.  Kilonovae are
promising EM counterparts because their emission is approximately isotropic
(compared to the beamed SGRB) and can peak at optical wavelengths, where
sensitive searches are possible \cite{Metzger&Berger12}.  Their brightness,
duration, and colors are diagnostics of physical processes during the merger.
Kilonovae also provide a unique probe to directly observe and quantify the
production of r-process nuclei.

The basic properties of kilonovae are understood with a toy model
\cite{Metzger+10}.  Approximate the ejecta at time $t$ after the merger as a
homogeneous sphere of mass $M$, uniform velocity $v$, radius $R \approx  v t$,
volume $V \approx 4\pi R^{3}/3$, and density $\rho = M/V$.  As the ejecta
expands, its thermal energy $E$ evolves according to
\be
\frac{dE}{dt} = \dot{E}_{\rm nuc} -L - P\frac{dV}{dt} = \dot{E}_{\rm nuc} -\frac{E}{t_{\rm diff}} - \frac{E}{t},
\label{eq:dEdt}
\ee
where the first term on the right hand side accounts for radioactive heating and 
the second term for radiation losses,
where $t_{\rm diff} = (3\kappa M)/(4\pi c v t)$ is the diffusion time 
and $\kappa$ is the mean opacity.
The last term in equation (\ref{eq:dEdt}) accounts
for adiabatic losses, where $P = E/(3V)$ is the pressure in the
radiation-dominated ejecta.  
The ejecta density is such that radiation can freely
escape from the bulk only once the expansion time $t = R/v$ equals $t_{\rm
diff}$, i.e. after a `peak' time
\be
t_{\rm pk} = \left(\frac{3\kappa M}{4\pi c v}\right)^{1/2} \approx 2.7\,{\rm day}\,\left(\frac{M}{10^{-2}M_{\odot}}\right)^{1/2}\left(\frac{v}{0.1\,\rm c}\right)^{-1/2}\left(\frac{\kappa}{\rm cm^{2}\,g^{-1}}\right)^{1/2}
\ee 

The opacity is dominated by 
Doppler-broadened atomic line
(bound-bound) transitions.  If the ejecta contains lanthanide or actinide
nuclei ($A \gtrsim 145$), then the optical opacity is very high $\kappa \gtrsim
10-100$ cm$^{2}$ g$^{-1}$
due to the complex atomic structure of the
f-shell valence electrons of these elements, resulting in a dense forest of
lines at optical/UV wavelengths \cite{Kasen+13,Tanaka&Hotokezaka13}.  On the
other hand, ejecta containing only lighter $r$-process elements ($A \lesssim
145$) with d-shell valence electrons will possess a lower opacity, $\kappa
\gtrsim 1$ cm$^{2}$ g$^{-1}$.  Experimental data is unfortunately not available
on most of the required line transitions, while theoretical atomic structure
models solving the many-body problem are statistical in nature.  The wavelength-
and composition-dependent opacity of partially ionized r-process nuclei remains
among the biggest uncertainties in modeling kilonovae.

For values of $M_{\rm ej} \sim 10^{-2}M_{\odot}$ and $v \sim 0.1$ c
characteristic of the merger ejecta (Fig.~\ref{f:just_combined_abundances}), we
find that $t_{\rm pk} \approx 1-10$ days, depending on $\kappa$.  The peak
luminosity 
is determined by the radioactive heating rate,
$\dot{E}_{\rm nuc} = M\dot{\epsilon}_{\rm nuc}$, on timescales of $t \sim
t_{\rm pk}$, where $\dot{\epsilon}_{\rm nuc}$ is the specific radioactive
heating rate.  The time dependence is a power-law 
due to overlapping contributions
of many decaying nuclei.  Radioactive energy is released in three channels \cite{Metzger+10}: (a)
the kinetic energy of beta decay electrons and fission fragments, (b)
neutrinos, (c) gamma-rays.  Although particle kinetic energy can be
efficiently shared with the plasma through Coulomb collisions, neutrinos escape
from the system and contribute no heating.  Gamma-rays may be trapped and
deposit their energy at early times $t \ll t_{\rm pk}$,
but are likely to escape at later times due to their lower opacity.  On
timescales of days after the merger, $\dot{\epsilon}_{\rm nuc} \approx
\epsilon_{\rm th}10^{10}(t/{\rm day})^{-1.3}$ erg s$^{-1}$ g$^{-1}$, relatively
independent of $Y_e \sim 0.1-0.3$ [e.g., \cite{Metzger+10,Roberts+11}], where
$\epsilon_{\rm th}<1$ accounts for the thermalization efficiency.  The peak luminosity of the kilonova is thus approximately given by 
\be
L_{\rm pk} \approx M\dot{\epsilon}_{\rm nuc}(t_{\rm pk}) \approx 5\times 10^{40}\,{\rm erg\,s^{-1}}\epsilon_{\rm th}\left(\frac{M}{10^{-2}M_{\odot}}\right)^{0.35}\left(\frac{v}{0.1\,\rm c}\right)^{0.65}\left(\frac{\kappa}{\rm cm^{2}\,g^{-1}}\right)^{-0.65},
\ee
and the effective temperature of the emission near its peak by
\be
T_{\rm pk} = \left(\frac{L_{\rm pk}}{4\pi \sigma R_{\rm pk}^{2}}\right)^{1/4} \approx 3460{\rm K}\epsilon_{\rm th}^{1/4}\left(\frac{M}{10^{-2}M_{\odot}}\right)^{-0.17}\left(\frac{v}{0.1\,\rm c}\right)^{-0.09}\left(\frac{\kappa}{\rm cm^{2}\,g^{-1}}\right)^{-0.41},
\label{eq:Tpeak}
\ee
where $R_{\rm pk} = v t_{\rm pk}$.  For $\kappa \gtrsim 10$ cm$^{2}$ g$^{-1}$,
characteristic of lanthanide-rich ejecta, the low temperature of $T_{\rm pk}
\sim 1000$ K implies a spectral peak in the near-IR
\cite{Barnes&Kasen13}. 

A candidate kilonova was discovered following the SGRB 130603B
\cite{Tanvir+13,Berger+13} 
based on the discovery 10 days after the burst of 1$\mu m$ emission in excess of that
predicted by an extrapolation of the power-law synchrotron afterglow emission.
Reproducing the peak luminosity of this event requires the ejection of $\sim
0.02-0.1M_{\odot}$ of heavy r-process nuclei.  A
clearer confirmation of kilonova emission in future events would require
obtaining a spectrum and identifying atomic absorption or emission lines from
r-process elements.  The velocity of the dynamical ejecta is sufficiently high
that lines will be Doppler-broadened, making them challenging to detect.
However, the slower disk outflows may give rise to detectable narrow lines
\cite{Kasen+15}. 

\subsubsection{Early Blue Emission}
\label{s:bluekilonova}

\begin{figure}
\includegraphics*[width=3in]{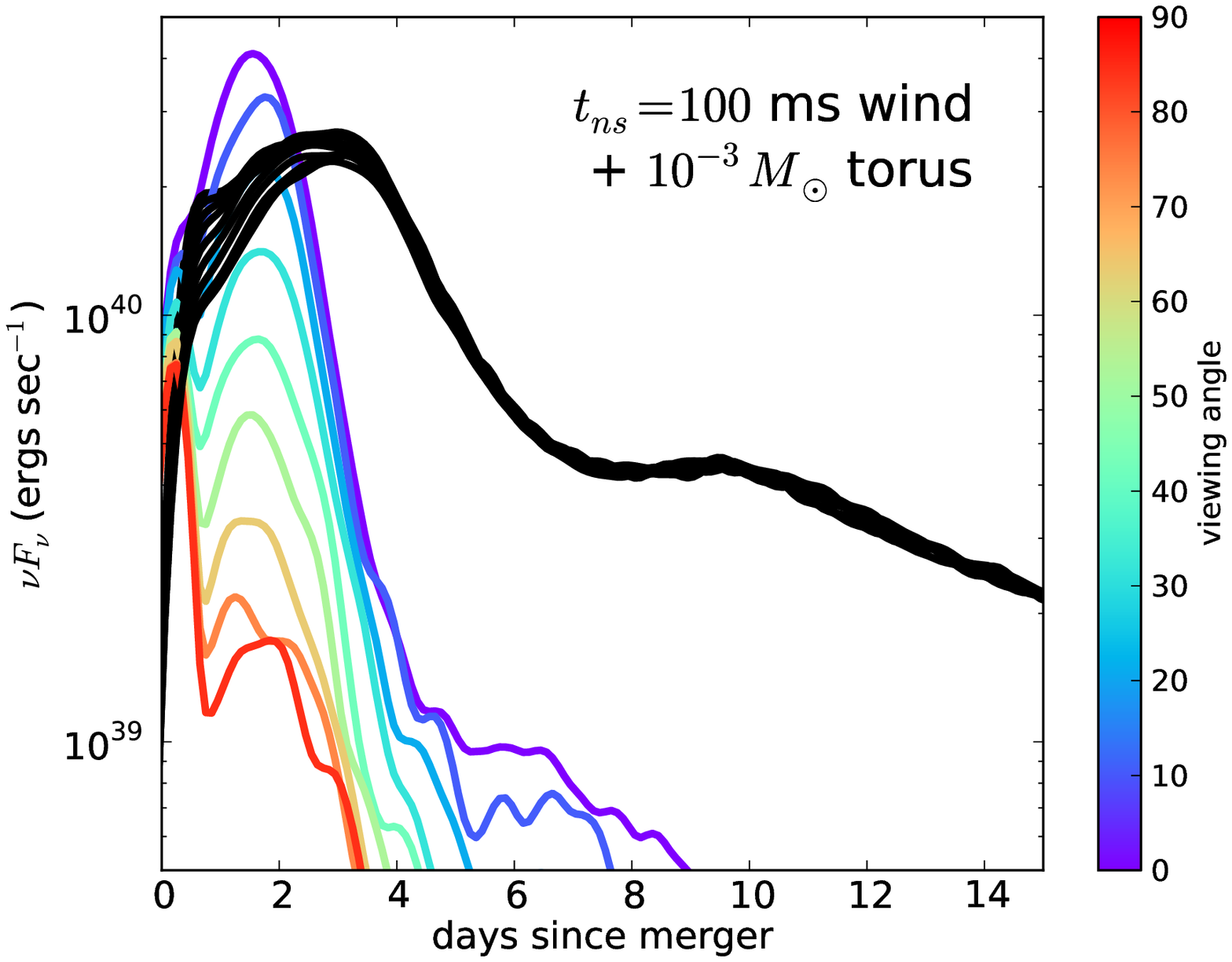} 
\caption{Kilonova light curves for a composite ejecta including
a lanthanide-poor disk wind from a HMNS remnant with lifetime $100$~ms, 
and a lanthanide-rich torus of dynamical ejecta.
Colored curves show the optical light curves as a function of observer
inclination from the rotation axis, while black curves show
near infrared emission. Figure from Kasen et al. \cite{Kasen+15}.}
\label{f:lightcurves_geometry}
\end{figure}

The low effective temperatures of kilonovae with lanthanide-rich ejecta are a
mixed blessing for GW follow-up.  On one hand, their extremely red colors
render them easily distinguishable from other astrophysical sources like
supernovae.  On the other hand, their discovery in the first place becomes much
more challenging because most EM follow-up facilities are sensitive at optical
wavelengths. The accretion disk outflows ($\S\ref{s:viscously_driven_wind}$) or shock-heated
dynamical ejecta \cite{wanajo2014}, in the case of a long-lived HMNS, 
can have a higher electron fraction and be
in some cases lanthanide-free.  
This could in
principle produce both an early `blue' (visual wavelength) phase of emission
lasting $\sim 1$ day, as well as a later `red' (infrared wavelength) phase
lasting $\sim 1$ week.  The relative fluxes of the blue and red emission
components would depend on the lifetime of the HMNS prior to collapsing into a
black hole \cite{Metzger&Fernandez14}, as well as on the observer viewing angle
\cite{Kasen+15} ({\bf Figure~\ref{f:lightcurves_geometry}}).

\subsection{Fast Ejecta and the Neutron Precursor}

Most of the dynamical ejecta in NSNS mergers remains sufficiently dense during
its expansion that all neutrons are captured into nuclei during the r-process.
However, \cite{Bauswein+13} found that a few percent of the ejected mass,
originating from the collision interface, expands sufficiently rapidly that
neutrons avoid being captured into nuclei (r-process freeze-out).  

These fast expanding free neutrons can have a dramatic impact on the kilonova
emission because they reside in the outermost layers of the ejecta.  Their beta
decays power a luminous `precursor' to the main kilonova emission, which peaks
within hours following the merger  \cite{Kulkarni05,Metzger+15}.  The neutron precursor
may encode information on the properties of the merging binary (e.g. NSNS
versus NSBH) and the NS EOS.  Future work is necessary to assess the robustness
of the fast-expanding ejecta (thus far seen with only one numerical code) and
to assess the survival of the neutrons in the face of leptonizing weak
interactions.  The fast expanding matter could also give rise to a broadband
non-thermal signature via its shock interaction with the ambient medium
\cite{Kyutoku+14}.

\subsection{Non-Radioactive Energy Sources}

Kilonovae can in principle be powered by more than just radioactivity, making
them brighter and easier to detect.  
One possible heating source is
shocks within the ejecta, for example if a late-time outflow from
the central engine were to collide with the expanding ejecta [\cite{Kisaka+15},
$\S\ref{s:EE}$].  A long-lived or stable NS remnant could also deposit its
rotational energy behind the ejecta via magnetic dipole radiation
(eq.~\ref{eq:LP}), forming the equivalent of a young pulsar wind nebula
[e.g., \cite{Yu+13,Metzger&Piro13,Siegel&Ciolfi15}].  This additional power can enhance
the kilonova luminosity by up to several orders of magnitude, if the NS remnant
possesses a dipole magnetic field of $B_{\rm d} \sim 10^{14}-10^{15}$ G. 

\subsection{Radio Transient from Ejecta Interaction with the Interstellar Medium}

Matter ejected during the merger slows down as it sweeps up gas in the
surrounding interstellar medium.  This deceleration is mediated by the forward
shock, which accelerates electrons to relativistic velocities and produces
non-thermal radio synchrotron emission \cite{Nakar&Piran11}.  

The ejecta transfer their energy to the ambient medium of density $n$ at the
characteristic radius ($R_{\rm dec}$) at which point they have swept up a mass
comparable to their own rest mass, $R_{\rm dec} \approx (3E_K/2\pi n m_p
v^{2})^{1/3}$, where $E_K= Mv^{2}/2 \approx 10^{50}(M/0.01M_{\odot})(v/0.1
c)^{2}$ erg is the ejecta kinetic energy.  This occurs at the deceleration
timescale,
\begin{equation}
  t_{\rm dec} \approx R_{\rm dec}/v \approx
  2.5\,{\rm yr}\,\,\left(\frac{E_{K}}{10^{50}\,{\rm erg}}\right)^{1/3}\left(\frac{n}{{\rm cm^{-3}}}\right)^{-1/3}\left(\frac{v}{0.3c}\right)^{-5/3},
\label{eq:tdec}
\end{equation} 
which sets the peak timescale of the radio emission.  If the observing
frequency $\nu$ is located above both the synchrotron peak frequency
and the self-absorption frequency, then the peak brightness for a souce at
distance $D$ is achieved at $t_{\rm dec}$, and is given by
\cite{Nakar&Piran11}:
\begin{equation}
  F_{\nu,\rm dec} \approx 0.08\,{\rm mJy}\,\, \left(\frac{E_{K}}{10^{50}\,{\rm erg}}\right)\left(\frac{n}{{\rm cm^{-3}}}\right)^{0.83}\epsilon_{e,-1}^{1.3}\epsilon_{B,-2}^{0.83}\left(\frac{v}{0.3c}\right)^{2.3}
 \left(\frac{D}{200\,{\rm Mpc}}\right)^{-2}\left(\frac{\nu}{{\rm GHz}}\right)^{-0.65},
\label{eq:Fp}
\end{equation}
where we have made the standard assumption that electrons are accelerated near
the shock into a power-law energy distribution, $N(E)\propto E^{-p}$ with
$p=2.3$ above a minimum Lorentz factor of $\gamma_m \sim 2$, and that
$\epsilon_B = 0.01\,\epsilon_{B,-2}$ and $\epsilon_e= 0.1\, \epsilon_{e,-1}$
are the fractions of post-shock energy density in the magnetic field and the
power-law electrons, respectively.

Equation (\ref{eq:Fp}) shows that the radio brightness is greatest for high
ejecta velocity and high external densities.  A high value of $n \sim 0.1-1$
cm$^{-3}$ is expected if the merger occurs in the disk of its host galaxy, but
$n$ could be much lower if the merger occurs outside of its host due to the
natal NS birth kicks, or in a globular cluster.  A low external density also
increases the rise time to $\sim t_{\rm dec} \gtrsim$ decade, making it
challenging to uniquely associate the radio transient with a GW event.  


\begin{issues}[FUTURE ISSUES]
\begin{enumerate}
\item Precursor emission will be challenging to detect, due to its relatively
      low luminosity and short rise time.  A coherent burst of radio emission from
      the NS magnetosphere orbiting through the conducting companion, while among the
      most promising observationally, is also the most challenging to predict
      theoretically. 

\item The detection of one or more GW sources will help constrain the NSNS/NSBH
      merger rate, which in turn will constrain the contribution of mergers to the
      Galactic $r$-process sources. A GW detection can also prove the existence of
      NSBH binaries, none of which has been  directly observed yet.

\item $r$-process nucleosynthesis predictions, which at present
      rely mostly on theoretically computed nuclear properties, will
      greatly benefit from new rare isotope facilities such as FRIB.
      Conversely, better theoretical models for the various ejecta components
      (e.g., tidal versus shock-driven dynamical ejecta, neutrino versus
      viscously-driven disk wind ejecta) can direct study towards relevant regions of
      parameter space to be probed by these experiments.

\item One of the biggest uncertainties in current kilonova models is the
      optical opacity of $r$-process elements in low ionization states, in particular
      of the Lanthanides and Actinides, for which experimental data is extremely
      limited.  Improvements in this area would enable more accurate predictions for
      the color and spectral line signatures of kilonovae.

\item The ability to detect the kilonova after a GW burst is sensitive to the
      presence of visual wavelength emission hours to days after the merger, which in
      turn depends on the electron fraction and geometric distribution of the ejecta.
      Detection and follow-up capabilities will greatly improve as more optical
      transient surveys, such as LSST, become operational.  Sensitive near-IR
      telescopes are needed for follow-up confirmation and characterization.
      Kilonova spectroscopy represents a promising target for planned thirty meter class
      telescopes.

\item Numerical simulations of NSNS/NSBH mergers 
      will continue to improve their physical 
      realism. Dynamical simulations need to combine MHD and neutrino
      transport with full GR to provide accurate estimates of the
      time to BH formation,
      the amount of mass dynamically ejected, and the mass in the
      accretion disk, in addition to better GW waveforms.
      The possible onset of a GRB jet is
      another major research direction.
      Post-merger models need to improve the treatment of
      angular momentum transport, including MHD, and the treatment
      of neutrinos, to improve the inputs for nucleosynthesis
      calculations and the computation of the EM signal
 
\end{enumerate}
\end{issues}


\section*{ACKNOWLEDGMENTS}
We thank Francois Foucart and Edo Berger for helpful comments on the manuscript.
RF acknowledges support from the University of California
Office of the President, and from NSF grant AST-1206097.
BDM gratefully acknowledges support from NASA Fermi grant NNX14AQ68G, NSF grant AST-1410950, NASA ATP grant NNX16AB30G, and the Alfred P. Sloan Foundation.

\bibliography{ms,rodrigo}
\bibliographystyle{ar-style5}

\end{document}